\documentclass[showpacs,preprintnumbers,amsmath,aps,amssymb]{revtex4-1}
\usepackage{graphicx}
\usepackage{dcolumn}
\usepackage{bm}
\usepackage{color}

\usepackage{soul}

\begin{document}

\title{On the Entropy of Deformed Phase Space Black Hole and  the Cosmological Constant}
 \author{A. Crespo-Hern\'andez}
\author{E. A. Mena-Barboza}
 \email{emena@cuci.udg.mx}
\affiliation{ Centro Universitario de la Ci\'enega, Universidad de Guadalajara\\
Ave. Universidad 1115 Edif. B, C.P. 47820 Ocotl\'an, Jalisco, M\'exico}%
\author{M. Sabido}
\email{msabido@fisica.ugto.mx}
\affiliation{ Departamento  de F\'{\i}sica de la Universidad de Guanajuato,\\
 A.P. E-143, C.P. 37150, Le\'on, Guanajuato, M\'exico
 }%

\date{\today}
\begin{abstract}
In this paper we study the effects of  noncommutative phase space deformations on the Schwarzschild black hole. This idea has been previously studied in {Friedmann-Robertson-Walker (FRW)} 
 cosmology, where this ``noncommutativity'' provides a simple mechanism that can explain the origin of the cosmological constant. In this paper we obtain the same relationship between the cosmological constant and the deformation parameter that appears in deformed phase space cosmology, but in the context of the deformed phase space black holes. This was achieved by comparing the entropy of the deformed Schwarzschild black hole with the entropy of the Schwarzschild-de Sitter  black hole.
 \end{abstract}
\keywords{black hole entropy; deformed phase space; cosmological constant}
 \maketitle
\section{Introduction}
The cosmological constant problem has been addressed  for several years and  remains as one of the central issues  in physics \cite{polchinski}. The  discovery  of the acceleration of the universe is usually attributed to a small, non-vanishing cosmological constant $\Lambda$. The different contributions to the vacuum energy density, from ordinary particle physics gives a value for $\langle\rho\rangle$ of order $M_p^4$  and should be canceled by the bare value of $\Lambda$. This cancellation  has to be better than $10^{-120}$, if we compare the zero-point energy of a scalar field  with the experimental value of $\langle\rho_{obs}\rangle$.  This incredible degree of fine tuning suggests that we are missing important physics \cite{cliff}. It  is likely  that the correct way to interpret the tiny value of the cosmological constant by conventional quantum field theory is not the whole story. This has lead many authors to suggest that the solution will come from an unconventional approach in fundamental physics. 
 Recently there has been mounting evidence of a late time accelerating universe, when one considers deformed phase space models \cite{vakili,obregon1,tachy1,maleko}. One is tempted to seriously consider the idea of noncommutative space-time to study the cosmological constant \cite{eri}. 
 
 The idea of a noncommutative space-time is not new, it has been studied from the  physics perspective by Snyder  \cite{wigner,snyder} and formal mathematical perspective was given by Connes \cite{connes}, during the last decades a lot of work has been done in noncommutative physics  \cite{nekra,connes_2,douglas, nc_grav,nc_grav2,nc_grav3,nc_grav4,nc_grav5,nc_grav6,connes2}. 
The noncommutative relationship between the space-time coordinates $[\hat{x}^i,\hat{x}^j]=i\Theta^{ij}$, gives an uncertainty principle between the coordinates that can be interpreted as length limit. Furthermore, noncommutativity is expected to be relevant at the Planck scale where it is known that usual semi classical considerations breakdown. 

From these ideas  noncommutative theories of gravity were proposed \cite{nc_grav,nc_grav2,nc_grav3,nc_grav4,nc_grav5,nc_grav6}. All of these formulations showed that the end result of a noncommutative theory of gravity, is a highly nonlinear theory. To avoid the difficulties of working with these highly nonlinear theories, and  to study the effects of noncommutativity on different aspects of the universe, noncommutative cosmology was proposed  \cite{ncqc}. The authors noticed   that the noncommutative deformations modify the noncommutative fields and  conjectured that the effects of the  full  noncommutative theory of gravity should be reflected in the minisuperspace variables.
 This was achieved  by introducing the Moyal product of functions in the Wheeler--DeWitt (WDW) equation of the Kantowski--Sachs (KS) cosmological model, in the same manner as is done in noncommutative quantum mechanics. This approach to noncommutativity has been used in connection  to the Hierarchy problem and to the cosmological constant problem  \cite{darabi, eri}. In the last five years, more general minisuperspace deformations have been used in cosmology. In several  models, the authors find an explicit relationship between  the cosmological constant $\Lambda$ and the deformation parameter  $\eta$ and  conjecture that the deformation parameter plays the role of a cosmological constant \cite{vakili,tachy1,maleko}. They arrive to this conclusion by comparing the scale factor of the deformed (``noncommutative'') model with the scale factor of a commutative model that includes $\Lambda$ and show that the deformation parameter plays the role of the cosmological constant  of the commutative model. 
 
 To further explore this relationship we consider a different gravitational system, the black hole.  Black holes are the perfect systems to study fundamental problems in gravity and therefore natural candidates to further explore the relationship between the cosmological constant and noncommutivity. In our approach we intend to study the origin of the cosmological constant and show that minisuperspace deformations are a feasible mechanism that can be responsible for its~origin. 

The main goal of this paper is to see if  a relationship between the deformation parameter $\eta$ and the cosmological constant $\Lambda$ arises in the context of deformed phase space black holes. We will introduce noncommutativity  in the phase space constructed from the minisuperspace variables,  by modifying the symplectic structure (Poisson's algebra of the minisuperspace)  \cite{darabi, ncqc, vakili,tachy1,maleko}. To calculate the entropy of the deformed black hole, we follow the procedure in \cite{ncbh, bertolami}, where the authors study  noncommutativity in black holes. We start with the phase space deformation \cite{darabi,vakili,tachy1,maleko}, in the KS cosmological model \cite{ncbh, piratas}.  Interchanging the coordinates $t \leftrightarrow r$,  we go  from a cosmological  solution to a black hole solution \cite{ncbh, bertolami}  and find the WDW equation for the noncommutative Schwarzschild black hole. As we are interested in obtaining the modification to the entropy of the black hole due to  the phase space deformation . We use the Feynman--Hibbs path integrals formalism \cite{hibbs} to calculate the thermodynamical properties of the black hole \cite{ncbh,tkach}. In the spirit of \cite{vakili,tachy1,maleko}, we compare the entropy of the deformed model with the entropy of a commutative model (Schwarzschild-de Sitter black hole) to find a relationship between $\Lambda$ and the deformation parameters.

The paper is organized as follows, in Section \ref{s2}, we present the model and introduce the noncommutative deformation. In Section \ref{s3}, we will show that, in our approach the origin of the cosmological constant and the deformation parameters are related. Section  \ref{s4},  is devoted for final~remarks.

\section{Deformed Black Hole Model}\label{s2}

In this section we present the model for the deformed phase space black hole. It is based on the relationship between the cosmological
KS metric and the Schwarzschild metric \cite{ro}. The
Schwarzschild solution can be written as
\begin{equation}
ds^{2}=-\left(1-\frac{2m}{r}\right)dt^{2}+\left( 1-\frac{2m}%
{r}\right)^{-1}dr^{2}
+r^{2}\left(d\theta^{2}+\sin^{2}\theta d\varphi^{2}\right) .
\end{equation}
For the case $r<2m$, the $g_{tt}$ and $g_{rr}$ components of the metric change
 their sign and $\partial_{t}$ becomes a space-like vector. If we make the
coordinate transformation $t\leftrightarrow r$, we find%
\begin{equation}
ds^{2}=-\left(\frac{2m}{t}-1\right)^{-1}  dt^{2}+\left(  \frac{2m}%
{t}-1\right)dr^{2}
+t^{2}\left(d\theta^{2}+\sin^{2}\theta d\varphi^{2}\right),
\end{equation}
when compared with the parametrization by Misner of the KS metric%
\begin{equation}
ds^{2}=-N^{2}dt^{2}+e^{\left( 2\sqrt{3}\gamma\right)} dr^{2}+
e^{\left(-2\sqrt{3}\gamma\right)}
 e^{\left(  -2\sqrt{3}\Omega\right)}  \left(
d\theta^{2}+\sin^{2}\theta d\varphi^{2}\right),
\label{KSmetric}%
\end{equation}
we identify
\begin{equation}
e^{ -2\sqrt{3}\gamma}e^{
-2\sqrt{3}\Omega}=t^{2},\quad e^{2\sqrt{3}\gamma}
=\frac{2m}{t}-1,\quad N^{2}=\left(\frac{2m}{t}-1\right)^{-1}.
\label{dif}
\end{equation}
This metric with the identification of the $N, \gamma$ and $\Omega$
functions  is also a classical solution  to Einstein's
equations. The
WDW equation for the
KS metric, with some particular factor ordering, is%
\begin{equation}
\left[  -\frac{\partial^{2}}{\partial\Omega^{2}}+\frac{\partial^{2}}%
{\partial\gamma^{2}}+48e^{ -2\sqrt{3}\Omega } \right]
\psi(\Omega,\gamma)=0.\label{ks}
\end{equation}
In \cite{ro1}, the authors use the WDW equation as a quantum equation for the Schwarzschild black hole and used as the starting point to calculate the entropy of black holes \cite{tkach}. 

 Deformed phase space black holes, satisfy a deformed Poisson algebra.
To construct  the deformed algebra 
we will follow the approach in \cite{tachy1}.
We start with
 a transformation on the classical phase space variables ${\{\Omega, \gamma, P_\Omega, P_\gamma\}}$, that satisfies the usual Poisson algebra
\begin{eqnarray}
\hat{\Omega}\rightarrow\Omega-\frac{\theta}{2}{P}_{\gamma},\qquad && \hat{\gamma}\rightarrow{\gamma} +\frac{\theta}{2}{P}_{\Omega}, \nonumber\\
\hat{P}_\Omega\rightarrow{P}_{\Omega} +\frac{\eta}{2}{\gamma}, \qquad && \hat{P}_\gamma\rightarrow{P}_{\gamma}-\frac{\eta}{2}{\Omega}. \label{12}\end{eqnarray}
$\hat{\Omega}, \hat{\gamma}, \hat{P}_{\Omega}, \hat{P}_{\gamma}$ are the deformed phase space variables and $\theta,\eta$ are the deformation parameters.
 It is easy to show that the deformed  minisuperspace variables satisfy the new algebra
\begin{equation}
[\hat{\Omega},\hat{\gamma}]=i\theta, [\hat{\Omega},\quad\hat{P}_{\Omega}]=[\hat{\gamma},\quad\hat{P}_{\gamma}]=i+\sigma,\quad [\hat{P}_{\Omega},\hat{P}_{\gamma}]=i\eta, \label{11}
\end{equation}
where $\sigma=\theta\eta/4$. Now that we have introduced the deformed phase space, we use the hamiltonian constraint for the KS model, but constructed with the variables that satisfy the modified algebra. When working with phase space deformations, two physical description arise. One is based on the original variables $\Omega$, $\gamma$ and another based on the new variables $\hat{\Omega}$, $\hat{\gamma}$. The first description has the interpretation  of a commutative theory where the effects of the deformation are encoded in a modified interaction, this description is referred as the ``{\it C-frame}'' interpretation. The second theory privilege the deformed variables, this is a theory with noncommutative variables but with the  original interaction and is usually referred as the ``{\it {NC}-frame}'' 
formulation. In this paper we will work in the ``{\it C-frame}'', therefore we can use a commutative space where the noncommutative effects are encoded in a modified potential \cite{barbosa_1,tachy2}. This allows us to use  the Feynman-Hibbs  method to calculate the entropy of the deformed phase space black hole.

With the transformations (\ref{12}), the WDW equation is deformed  and exhibits an explicit dependence on the noncommutative parameters
\begin{equation}
\left[-\left(-i\frac{\partial}{\partial \Omega}+\frac{\eta}{2}\gamma\right)^2+\left(-i\frac{\partial}{\partial \gamma}-\frac{\eta}{2}\Omega\right)^2
 -48e^{-2\sqrt{3}\left[\Omega+\frac{i\theta}{2}\frac{\partial}{\partial \gamma}\right]}\right]\psi(\Omega,\gamma)=0.\label{14}
\end{equation}
We use the anzats $\psi_\nu(\Omega,\gamma)=\Xi(\Omega)e^{\left[i\left(\nu-\frac{\eta}{2}\Omega\right)\right]\gamma}$
in Equation (\ref{14}) to obtain
\begin{equation}
-\Xi''(\Omega)-(\eta\Omega-\nu)^2\Xi(\Omega)
+48\text{exp}\left[-2\sqrt{3}(1+\sigma)\Omega+\sqrt{3}\theta \nu\right]\Xi(\Omega)=0.\label{20}
\end{equation}
This second order ordinary differential equation can be solved numerically and the solutions depend  on the deformation parameters  $\theta$ and $\eta$.  For our purposes we don't need analytical solutions. To compute the partition function of the deformed black hole, we use the Feynman--Hibbs procedure~\cite{hibbs}. 
 This procedure is based in exploiting the similarities of the expression of the density matrix and the kernel of Feynman's path integral approach to quantum mechanics. By doing a Wick rotation $t\to i\beta$ we get the Boltzmann factor and the  kernel is transformed to the density matrix. 
 { The kernel is calculated along  the paths that go from $x_1$ to $x_2$, if we consider small $\Delta t$ (small $\beta$). Then when calculating the partition function, only the paths that stay near $x_1$ have a non negligible contribution (the exponential in the expression for the density matrix gives a negligible contribution to the sum from the other paths). Therefore, the potential to a first order approximation can be written as $V(x)\approx V(x_1)$, for all the contributing paths.
 In this approximation, }we can formally establish a map from the path integral formulation of quantum mechanics to the classical canonical partition function. To introduce quantum mechanical effects, we { must incorporate the changes to the potential along the path}, in particular we are interested in the first order effects. For this we start by doing a Taylor expansion around the mean position $\tilde x$ along any path. Calculating the kernel with $\tilde x$ and doing the Wick rotation, we get the modified partition function. { This partition function is calculated in a classical manner but with the corrected potential, and the quantum effects are encoded in the corrected potential (the potential calculated along the mean value of the path).  The effective potential is a mean value of the potential $V(x)$ averaged over points near $\tilde x$ with a gaussian distribution. Therefore to incorporate the quantum effects we only need to calculate the corrected potential}, and simply calculate the partition function using the corrected potential \cite{ncbh,tkach}. 
Because we are using the ``{\it C-frame}'' interpretation for the deformed phase space model, we can assume  that we are working with commutative variables and  the noncommutative effects are encoded in the potential
\begin{equation}
V(\Omega)=48e^{\left[-2\sqrt{3}(1+\sigma)\Omega + \sqrt{3}\theta \nu\right]}-(\eta\Omega-\nu)^2.
\end{equation}
Furthermore, to calculate the canonical partition function we can use the Feynman--Hibbs approach to the noncommutative potential \cite{tkach,ncbh,bertolami}.
We start by expanding the  potential   to second order in $\Omega$ and making the change of variable $\Omega\rightarrow \frac{1}{\sqrt{6}}\chi+b$ where  $b$ is a constant
\begin{equation}
b=\frac{a\eta-48\sqrt{3}(1+\sigma)e^{\sqrt{3}\theta a}}{\eta^2-288(1+\sigma)^2e^{\sqrt{3}\theta a}}.
\end{equation}
Now multiplying by  Planck{'}s energy ${E_p}$ and setting $\chi=\frac{x}{l_p}$ Equation (\ref{20}) takes the form
\begin{eqnarray}
&&-\frac{l_p^2E_p}{2}\Xi''(x)+\left(4(1+\sigma)^2e^{\sqrt{3}\theta \nu}-\frac{\eta^2}{72}\right)\frac{E_p}{l_p^2}x^2\Xi(x)\\
&&=\frac{E_p}{12}\left[\nu^2+\frac{\left(\nu\eta -48\sqrt{3}(1+\sigma)e^{\sqrt{3}\theta \nu}\right)^2}{288(1+\sigma)^2e^{\sqrt{3}\theta \nu}-\eta^2}-48e^{\sqrt{3}\theta \nu}\right]\Xi(x). \nonumber\label{27}
\end{eqnarray}
The Feynman-Hibbs procedure allows  to incorporate the quantum corrections to the partition function through the corrected potential, which results in 
\begin{equation}
U(x)=\frac{3}{4\pi}\left((1+\sigma)^2e^{\sqrt{3}\theta \nu}-\frac{\eta^2}{288}\right)\frac{E_p}{l_p^2}\left[x^2+\frac{\beta l_p^2E_p}{12}\right],
\end{equation}
the corrected partition function is 
\begin{equation}
Z=\sqrt{\frac{2\pi}{3}}\left((1+\sigma)^2e^{\sqrt{3}\theta \nu}-\frac{\eta^2}{288}\right)^{-\frac{1}{2}}
\frac{e^{-\frac{1}{16\pi}\left((1+\sigma)^2e^{\sqrt{3}\theta \nu}-\frac{\eta^2}{288}\right)\beta^2E_p^2}}{\beta E_p}, \label{38}
\end{equation}
from which we can proceed to calculate thermodynamic properties.

The internal energy  is $\langle E\rangle=-\frac{\partial}{\partial\beta}\ln Z$:
\begin{equation}
\langle E\rangle=\frac{1}{\beta}+\frac{1}{8\pi}\left((1+\sigma)^2e^{\sqrt{3}\theta \nu}-\frac{\eta^2}{288}\right)\beta E_p^2. \label{39}
\end{equation}
For the black hole  $\langle E\rangle=mc^2$ and solving the quadratic equation for $\beta$ we get the temperature of the deformed phase space black hole. The temperature of the black hole can be written in terms of the deformation parameters and mass of the black hole and defining the ``noncommutative'' temperature~$\hat{\beta}_H$
\begin{equation}
\hat{\beta}_H=\frac{8\pi mc^2}{\left((1+\sigma)^2e^{\sqrt{3}\theta \nu}-\frac{\eta^2}{288}\right)E_p^2},\label{temp}
\end{equation}
then we can write a simple expression for the deformed phase space black hole temperature.
\begin{equation}
\beta=\hat{\beta}_H\left(1-\frac{1}{mc^2\hat{\beta}_H}\right).
\end{equation}
It is easy to  verify that for the appropriate  values of the deformation parameter reproduce results in \cite{tkach,ncbh,kastrup}. 
 { Usually the temperature is defined by the elimination of conical singularities at the horizon of the Euclidean black hole geometry. In order to determine the temperature  of the horizon of the deformed black hole, we need to take the full noncomutative gravity action (which is a highly no linear  \cite{nc_grav,nc_grav2,nc_grav3,nc_grav4,nc_grav5,nc_grav6}), find the deformed black hole metric  and calculate the temperature by resolving the conical singularity. Due to the complexity of  noncommutative theories of gravity, we choose to follow the approach of noncommutative cosmology: introduce noncommutativity in the minisuperspace. Because we are using a well defined formalism to calculate the thermodynamics of the black hole that in the commutative case gives the correct temperature  \cite{tkach,ncbh,kastrup}. Therefore, we can be confident that the calculated temperature for the deformed phase space black hole, is well defined in the context of the Feynman--Hibbs approach, but we can not prove that it is equivalent to the removal of a conical~singularity.}

The entropy is calculated from $S=k\beta\langle E\rangle+k\ln Z,$ and find  the deformed Hawking--Bekenstein entropy as 
\begin{eqnarray}
\frac{S}{k}=\frac{\hat{S}_{BH}}{k}-\frac{1}{2}\ln\left[\frac{\hat{S}_{BH}}{k}\right]+\mathcal{O}\left[\left(\frac{\hat{S}_{BH}}{k}\right)^{-1}\right], \label{endagts1}
\end{eqnarray}
where
\begin{equation}
\frac{\hat{S}_{BH}}{k}=\frac{4\pi m^2c^4}{\left((1+\sigma)^2e^{\sqrt{3}\theta \nu}-\frac{\eta^2}{288}\right)E_p^2},\nonumber
\end{equation}
as with the temperature,  for  $\theta=0$ and $\eta=0$,  we recover the commutative result. For $\eta=0$ and $\theta \ne 0$, we  obtain the results  in \cite{ncbh}.
Finally, when the deformation is only present  on the canonical momentum  ($\theta=0$ and $\eta\ne 0$),   with  $E_p=c=k=1$, we get 
\begin{equation}
S_{BH}^{\eta}=4\pi m^2\left(1+\frac{\eta^2}{288}+\dots\right),\label{nc}
\end{equation}
to first order in the  $(\eta)^2$. 

As expected the phase space deformation gives a correction  to the Schwarzschild black hole~entropy.



\section{Deformed Phase Space Parameters and $\Lambda$}\label{s3}
During the last few years mounting evidence on the relationship between $\Lambda$ and the deformation parameter $\beta$ has appeared \cite{vakili,tachy1,maleko}. To find an expression for $\Lambda_{eff}$,  the authors compare the de Sitter cosmology scale factor with the scale factor of the deformed phase space model in the limit $t\to \infty$. In this limit, the scale factor  of the deformed model  behaves as the de Sitter scale factor and an effective cosmological constant $\Lambda_{eff}$  can be defined. This effective cosmological constant is a function of  the deformation parameters.  In order to obtain the relationship between the deformation parameters and the cosmological constant, we will compare a commutative model that includes $\Lambda$ with a deformed phase space model that does not include $\Lambda$.
To establish the connection between $\Lambda$ and the parameter $\eta$,  we choose as the commutative  gravitational system with a cosmological constant, the Schwarzschild-de Sitter black hole.

The metric for the  Schwarzschild-de Sitter black hole model is
\begin{equation}
ds^2=-\left(1-\frac{2m}{r}+\frac{\Lambda r^2}{3}\right)dt^2
+\frac{dr^2}{ 1-\frac{2m}{r}+\frac{\Lambda r^2}{3}}
+r^2\left(\text{d}\theta^2+\sin^2\theta\text{d}\phi^2\right),
\end{equation}
which it depends on the mass and the cosmological constant. There is a singularity at $r=0$ and two horizons. The smaller one  $r_b$ is called the black hole horizon and the larger one $r_c$ the cosmological horizon. As explained in \cite{teitelboim}, we have two options, either use $r_c$ or $r_b$ as boundaries.{\  When calculating thermodynamic properties, this can be physically interpreted  as  two horizons that are not in thermal equilibrium \cite{t2}.
For our purposes we will restrict to the black hole horizon. 

The black hole horizon radius can be written as 
\begin{equation}
r_b=\frac{2}{\sqrt{\Lambda}}\cos\left(\frac{\pi+\chi}{3}\right),\label{rh}
\end{equation}
where $\chi=\cos^{-1}(3\sqrt{\Lambda m^2})$.

{ The mass  can be defined  in terms of the black hole horizon  \cite{bousso, julio}
\begin{equation}
m=\frac{r_b}{2}\left(1-\frac{\Lambda r^2}{3}\right),
\end{equation}
$m$ is only defined when $r$ takes values between the two horizons, for vanishing $\Lambda$  we get the value for Schwarzschild black hole.  

To calculate the entropy we follow as usual, thus the entropy is proportional to the area $S\sim \pi r_b^2$.
Expanding  the black hole  radius $r_b$ to third order in $\sqrt{\Lambda}m$, we get for the entropy}
\begin{equation}
S=
\frac{\pi r_b^2}{4}\approx\pi m^2\left(1+\frac{8\Lambda m^2}{3}\right)+\cdots.\label{21}
\end{equation}
As already stated, where interesting of comparing the deformed phase space  Schwarzschild black hole with the commutative Schwarzschild-de Sitter black hole.
Comparing (\ref{21})  with the entropy of the deformed model  (\ref{nc}) we find that  $\Lambda\sim\frac{\eta^2}{m^2}$.
 \section{Conclusions}\label{s4}
In this paper we have explored the possible relationship between the cosmological constant $\Lambda$ and the noncommutative deformation parameter of the canonical momentum. This relationship has been suggested when studying the late time behaviour of several cosmological models and comparing with the commutative de Sitter cosmology \cite{tachy1}. The results point out a quadratic dependence of the cosmological constant with the noncommutative parameter suggesting that the origin of the cosmological constant could be related to phase space deformations.

The approach we followed is similar, we compare  the deformed phase space  Schwarzschild black hole   with the commutative { Schwarzschild-}de Sitter black hole.  We take advantage of the relationship between the Schwarzschild metric and the KS cosmological model,   so that we can use the temperature and entropy of the deformed phase space black hole, calculated from the deformed KS-WDW equation~\cite{bertolami}.  Comparing  with the entropy of the  Schwarzschild-de Sitter black hole in the limit $\eta<<1$ and $\theta=0$, we find that $\Lambda\sim\frac{\eta^2}{m^2}$. This is consistent with the results obtained in the cosmological scenario \cite{tachy1}, where  a quadratic relationship between the cosmological constant and the deformation parameter has been obtained.
 It is encouraging  that we have obtained the same result  in a different gravitational system, but we can not say that the deformation parameter can always replace $\Lambda$.  We need to establish the relationship between $\Lambda$ and $\eta$  under less restrictive conditions.  This is under research  and will be reported elsewhere. 

\acknowledgments{This work is  supported by  CONAC yT grants 167335, 179208, 257919, 290649 and DAIP1107/2016 and by C.U.CI., U. de G. project Desarrollo de la investigaci\'on y fortalecimiento del posgrado 235506, A. Crespo is supported by CONACyT PhD. grant.}
\\{\bf Author contributions:} All of the authors conceived of, designed and performed the calculations together. In
addition, they analyzed the results and wrote the paper together. All authors have read and approved the final
manuscript.} The authors contributed to this work equally.\\

{\bf Conflict of interests:} The authors declare no conflict of interest. 
\renewcommand\bibname{References}

\end{document}